\documentclass[12pt]{article}

\usepackage{a4}
\usepackage{epsfig}

\begin{document}

\noindent\Large\bf
 Winner-Relaxing
 Self-Organizing Maps
\\\\
\\
\noindent\large\bf
Jens Christian Claussen\\
\small
\sl
 claussen\verb+@+theo-physik.uni-kiel.de\\
Institut f\"ur Theoretische Physik und Astrophysik,
Leibnizstr.\ 15,
\\
Christian-Albrechts-Universit\"at zu Kiel, 
24098 Kiel, Germany
\\
\\
\bf
A new family of self-organizing maps,
the Winner-Relaxing Kohonen Algorithm, 
is introduced as a generalization of a 
variant given by Kohonen in 1991.
The magnification behaviour is calculated 
analytically.
For the original variant a magnification 
exponent of 4/7 is derived; 
the generalized version allows to steer the
magnification in the wide range
from exponent 1/2 to 1 in the one-dimensional case,
thus  provides optimal mapping in the sense of
information theory. The Winner Relaxing Algorithm 
requires minimal extra computations
per learning step and is conveniently easy to implement.
\normalsize \rm
\section{Introduction}
The self-organizing map (SOM) algorithm
(Kohonen 1982)
served both as model
for topology-preserving primary sensory processing
in the cortex 
(Obermayer et al.\ 1992),
and for technical applications 
(Ritter et al.\ 1992).
Self-organizing feature maps map an input space, such as the
retina or skin receptor fields, into a neural layer by
feedforward structures with lateral inhibition.
Defining properties
are topology preservation, error tolerance,
plasticity,
and self-organized formation by a local process.
Compared to other clustering algorithms and
vector quantizers its apparent advantage
for data visualization and exploration
is its approximative topology preservation.
In contrast to the Elastic Net
(Durbin \& Willshaw 1987)
and the Linsker (1989) Algorithm,
which are performing gradient descent in a certain
energy landscape, the Kohonen algorithm
lacks
an energy function in the general case
of a continuous input distribution.
Although the learning process can be described in terms of
a Fokker-Planck equation (Ritter \& Schulten 1988), 
the expectation value of
the learning step is a nonconservative force 
(Obermayer et al.\ 1992)
driving the process so that it has no associated energy function.
Despite a lot of research, the relationships between the
Kohonen model and its variants to general principles
remain an open field  (Kohonen 1991).
\\\\
To appear in {\sl Neural Computation}
\clearpage
\subsection{Kohonen's Self Organizing Feature Map}
Kohonen's Self Organizing Map
is defined as follows:
Every stimulus ${\bf v}$ of an input space $V$
is mapped to 
a ``center of excitation'', or winner 
\begin{eqnarray}
\mbox{${\bf s}$}=
\mbox{$\mbox{\rm argmin}_{{\bf r}\in R}
|{\bf w}_{{\bf r}} -{\bf v}|$},
\label{eq:voronoi}
\end{eqnarray}
where $| . |$ denotes the Euclidian distance in input space.
In the Kohonen model, the learning rule for each synaptic
weight vector ${\bf w}_{{\bf r}}$ is given by
        \begin{eqnarray}
\delta {\bf w}_{{\bf r}} 
         = \eta \cdot g_{{\bf r} {\bf s}}
         \cdot ({\bf v}-{\bf w}_{{\bf r}}),
\label{som_learning}
        \end{eqnarray}
where $g_{{\bf r}{\bf s}}$ defines the neighborhood relation in $R$,
and will throughout this paper be 
a Gaussian function of the Euclidian
distance $|{\bf r}-{\bf s}|$ in the neural layer.
Topology preservation is enforced by the common
update of all weight vectors whose neuron ${\bf r}$ is adjacent
to the center of excitation ${\bf s}$;
the adjacency function $g_{{\bf r}{\bf s}}$ prescribes the topology in the
neural layer. 
The speed of learning  $\eta$ usually is decreased  during the process.

\subsection{\hspace*{-.4em}The Winner Relaxing Kohonen Algorithm}
\begin{figure}[htbp]
\centerline{\epsfig{file=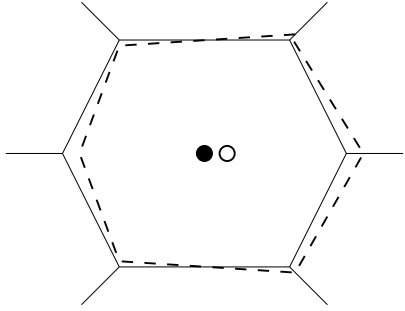,width=0.4\columnwidth}}
\caption{\small\sf
Shift of Voronoi borders as an effect of weight vector update.
\label{voronoishift}}
\end{figure}

We now consider an energy function $V$ first proposed in 
(Ritter et al.\ 1992).
If we have a discrete input space,
the potential function
for the expectation value of the learning step is given by
\begin{eqnarray}
V (\{{\bf w}\}) = \frac{1}{2} \sum_{{\bf r}{\bf s}}
g^{\gamma}_{{\bf r}{\bf s}}
\sum_{\mu|{\bf v}^{\mu}\in F_{{\bf s}}(\{{\bf w}\})}
p({\bf v}^{\mu})  \cdot  |{\bf v}^{\mu} -{\bf w}_{{\bf r}}|^2,
\label{eq:ritterpotential}
\end{eqnarray}
where $F_{{\bf s}}(\{{\bf w}\})$ is the cell of the {\em Voronoi
tesselation}  (or {\em Dirichlet tesselation})
of input space
defined by
(\ref{eq:voronoi}).
For discrete input space, where $p({\bf v})$ is a sum over
delta peaks $\delta({\bf v}-{\bf v}^{\mu})$,
the first derivative 
w.r.t.\ ${\bf w}_{{\bf r}}$
is not continuous at all weight vectors
where the borders of the voronoi tesselation are shifting over
one of the input vectors (Fig.~\ref{voronoishift}).
However, (\ref{eq:ritterpotential}) requires the assumption that
none of the borders of the Voronoi tesselation is shifting
over a pattern vector ${\bf v}^{\mu}$,
which may be fulfilled in the final convergence phase
for discrete input spaces, but becomes 
problematic if there are more receptor positions
than neurons.
If $p({\bf v})$ is continuous,
the sum over $\mu$ becomes an integral,
and with every stimulus vector update the
surrounding Voronoi borders slide over stimuli 
(which means they become represented by annother weight vector),
so that there is no global energy function 
for the general case.

We remark that replacing the crisp (or hard)
winner selection 
(\ref{eq:voronoi}) by
a soft-winner 
${\bf s}=\mbox{argmin}_{\bf r} \sum_{{\bf r'}} 
 g_{{\bf rr'}}|{\bf w}_{{\bf r'}}-{\bf v}|^2$
minimizes
 (\ref{eq:ritterpotential})
even in the continuous case
(Graepel et al 1997, Heskes 1999).
This is a formally elegant approach if one
wants to ensure the existence of an energy function
and accepts to modify the winner selection.

However, to motivate the Winner Relaxing learning,
we return to the hard winner selection scheme
(\ref{eq:voronoi}) and take up
the learning rule given by Kohonen (1991).
Our use of this ansatz however is justified here
only {\sl a posteriori} by its use for adjusting the magnification.

From the shift 
of the borders of the Voronoi tesselation
$F_{{\bf s}}(\{{\bf w}\})$ 
(see Fig.~\ref{voronoishift})
in evaluation of the gradient
with respect to a weight vector ${\bf w}_{{\bf r}}$,
Kohonen (1991) derived for the (approximated) 
gradient descent in $V$
the additive term 
$ -   \frac{1}{2}  \eta   \delta_{{\bf r}{\bf s}}
   \sum_{{\bf r}^{'}\neq{\bf s}}
   g_{{\bf r}^{'}{\bf s}}
   ({\bf v}-{\bf w}_{{\bf r}^{'}})$
extending (\ref{som_learning})
for the winning neuron.
As it implied an additional elastic relaxation,
it was straightforward to call it
`Winner Relaxing' (WR) Kohonen algorithm (Claussen 1992).
In the remainder we study the 
 (generalized) Winner Relaxing Kohonen algorithm,
or Winner Relaxing Self-Organizing Map (WRSOM),
introduced firstly in (Claussen 1992),
in the form
\begin{eqnarray}
\delta {\bf w}_{{\bf r}} &=& \eta\{({\bf v}-{\bf w}_{{\bf r}})
                 g^{\gamma}_{{\bf r}{\bf s}}
   - \lambda   \delta_{{\bf r}{\bf s}}
   \sum_{{\bf r}^{'}\neq{\bf s}}
      g^{\gamma}_{{\bf r}^{'}{\bf s}}
      ({\bf v}-{\bf w}_{{\bf r}^{'}})\},
\label{eq:wr_upd_ohnemu}
\end{eqnarray}
where ${\bf s}$ is the center of excitation for incoming stimulus
${\bf v}$, and $g^{\gamma}_{{\bf r}{\bf s}}$ is a Gaussian function
of distance in the neural layer with characteristic length $\gamma$.
Here $\lambda$ is a 
free parameter of the algorithm.
The original algorithm 
(associated with the potential function)
proposed by Kohonen in 1991 is
obtained for $\lambda=+1/2$, whereas the classical
Self Organizing Map Algorithm is obtained
for $\lambda=0$. 
The influence of $\lambda$ on the magnification behaviour 
is the central issue of this paper.

\clearpage
\subsection{The Magnification Factor}
The magnification factor is defined as the density of
neurons ${\bf r}$ 
(i.e.\ the density of
synaptic weight vectors ${\bf w}_{{\bf r}}$)
per unit volume of input space, and therefore is given by the
inverse Jacobi determinant of the mapping
from input space to neuron layer:
$M=|J|^{-1}=|\det({{d}}{\bf w}/{{d}}{\bf r})|^{-1}$.
We assume the input space to be continuous and of same dimension
as the neural layer,
and the map to be noninverting $(J>0)$.

The magnification factor
quantifies the networks' response
to a given probability density of stimuli $P({\bf v})$.
To evaluate $M$ in higher dimensions, one in general has to
compute the  equilibrium state of the whole network and
needs therefore the complete global knowledge on $P({\bf v})$,
except for separable cases.
For one-dimensional mappings
 the magnification factor can follow
an universal magnification law, that is,
$M(\bar{{\bf w}}({\bf r}))$ is a function of the local
probability density $P$ only, independent of both the location
${\bf r}$ in the neural layer and the location
$\bar{{\bf w}}({\bf r})$ in input space.
Hereby it is nontrivial whether there exists a power law
or not; the Elastic Net
obeys an universal 
magnification law that remarkably is not a
power law (Claussen \& Schuster 2002)
due to a nonvanishing elastic tension 
in regions of small input density.
For the classical Kohonen algorithm the magnification law
is given by a power law $M(\bar{{\bf w}}({\bf r}))
\propto P(\bar{{\bf w}}({\bf r}))^{\rho}$ with exponent
$\rho=\frac{2}{3}$ 
(Ritter \& Schulten 1986).
See Table~\ref{tab_magniexpo} 
for an overview.
For a discrete neural layer and 
different neighborhood kernels corrections apply
(Ritter 1991, Ritter et al.\ 1992, Dersch\& Tavan 1995).

\begin{table}[htbp]
\begin{center}
\begin{tabular}[t]{|c|c|c|c|c|c|}
\hline
Elastic  & VQ, & WRK  &SOM& WRK   & Linsker\\
Net & NG &  $\lambda=\frac{1}{2}$ &$\lambda=0$ &  $\lambda=-1$  & 
\\
\hline
&&&&& \\[-0.2ex]
{
 $\displaystyle\frac{1}{1+\frac{\kappa}{\sigma^2}\frac{P}{J}} $}
& $P^{\displaystyle 1/3}$
& $P^{\displaystyle 4/7}$
& $P^{\displaystyle 2/3}$
& $P^{\displaystyle 1}$
& $P^{\displaystyle 1}$
\\[0.1ex]
&&&&& \\[-0.3ex] 
\hline \end{tabular} \end{center} 
\vspace{-2ex}
\caption[Magnification laws]{\small\sf
Magnification laws
for one-dimensional maps
\label{tab_magniexpo}}
\end{table}

As the brain is assumed to be optimized by evolution 
for information processing, one could conjecture that 
maximal mutual information can define an extremal principle
governing the setup of neural structures.
For feedforward neural structures with lateral 
inhibition,  an algorithm of maximal mutual information
has been defined by Linsker (1989)
using the gradient descend in mutual information.
It requires computationally costly integrations,
and has a highly nonlocal learning rule;
therefore it is neither favourable 
as a model for biological maps,
nor feasible for technical applications.
Due to realization constraints, 
both technical applications and cortical networks
(Plumbley 1999)
are not necessarily
capable of reaching this optimum.
Even if one had
experimental 
data
of the magnification behaviour, the
question from what self-organizing dynamics 
neural structures emerge, remains.
Overall it is desirable to find 
learning rules
that minimize mutual information in a simpler way.

An optimal map from the view of information theory would
reproduce the input probability exactly
($M\sim P({\bf v})^{\rho}$ with $\rho=1$),
being equivalent to the
condition that all neurons in the layer are firing with same
probability.
This defines an equiprobabilistic mapping 
(van Hulle 2000). 
An exponent  \mbox{$\rho=0$}, on the other hand, corresponds to
a uniform distribution of weight vectors, 
or no adaptation at all.
So the magnification exponent is a direct indicator, how far a
Self Organizing Map algorithm is away from the optimum predicted
by information theory.

\section{\mbox{Magnification Exponent of the}
 \protect\linebreak[4] Winner-Relaxing Kohonen Algorithm}
We now derive the magnification law of the Winner-Relaxing
Kohonen algorithm (\ref{eq:wr_upd_ohnemu}) 
for the case of a 1D$\to$1D  map.
Note that for higher dimensions analytical results can only be obtained
for special degenerate cases of the input probability density and therefore
lack generality.

\begin{figure}[htbp]
\centerline{\epsfig{file=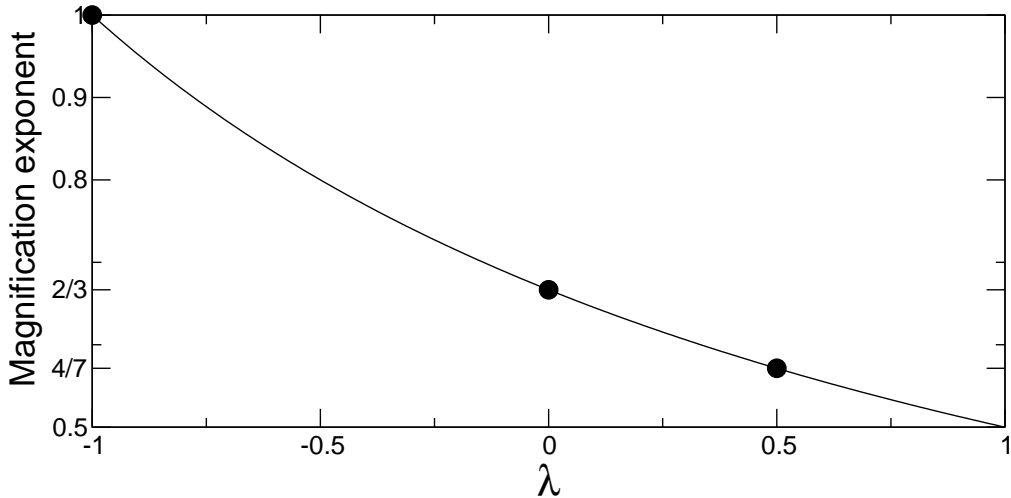,width=1.0\columnwidth}}
\caption{\small\sf
Impact of parameter $\lambda$ on the magnification
exponent. The cases of $\lambda=1/2$ (Kohonen 1991), the SOM 
case $\lambda=0$ (Kohonen 1982)
and the ``winner enhancing'' choice $\lambda=-1$ are marked with dots.
\label{fig:wrktheo}}
\end{figure}

The necessary condition for the final state
of the algorithm  is that
the expectation value of the learning step vanishes
for all neurons $r$:
\begin{eqnarray}
\forall_{{r}{\in}R} \;\;\;\; 0 =
\int{\rm d}{v}\;p({v})\delta\bar{w}_{{r}}({v}).
\label{eq:notwend}
\end{eqnarray}
Since this expectation value is equal to the learning step of
the pattern parallel rule,
(\ref{eq:notwend}) is the
stationary state condition for {\em both} serial and parallel
updating, and also for batch updating.
Thus we can proceed for these variants simultaneously
(As synaptic plasticity is widely assumed to be based on
integrative effects, one could claim that a parallel model
is sufficient).
The update rule (\ref{eq:wr_upd_ohnemu})
can be extended by an additional diagonal term controlled by 
$\mu$\footnote{
Whereas the extra term controlled by the parameter $\mu$ has
been introduced in (Claussen 1992) for pure generality,
and will be kept within the derivation,
it does not contribute to the magnification.
In general, the setting $\mu=0$, 
is recommended (and probably most stable), 
and the Winner Relaxing Kohonen algorithm
thus has only one relevant control parameter $\lambda$.}:
\begin{eqnarray}
\delta {w}_{{r}} &=& \eta \{ ({v}-{w}_{{r}})
                         \cdot  g^{\gamma}_{{r}{s}}
   + \mu  ({v}-{w}_{{r}})                \delta_{{r}{s}} 
\nonumber\\& &
   -   \lambda    \delta_{{r}{s}}
   \sum_{{r}^{'}\neq{s}}
   g^{\gamma}_{{r}^{'}{s}}
   ({v}-{w}_{{r}^{'}}) \}.\label{eq:wrkupd}
\end{eqnarray}
By insertion of the update rule
(\ref{eq:wrkupd}) one obtains
\begin{eqnarray}
0 &=& \int {\rm d}s \; P(\bar{w}(s)) J(s)  g^{\gamma}_{rs} (\bar{w}(s)-\bar{w}(r))
  \nonumber\\ & & 
+ (\mu+\lambda) \cdot\underbrace{\int {\rm d}s\;
 P(\bar{w}(s)) J(s) 
 \delta_{rs} (\bar{w}(s)-\bar{w}(r))
              }_{\equiv 0}       \nonumber\\
& & \;\;  -  \lambda \cdot \int \int {\rm d}s \; {\rm d}r^{'} \;
   P(\bar{w}(s)) J(s)   \delta_{rs}  g^{\gamma}_{r^{'}s}
          (\bar{w}(s)-\bar{w}(r^{'}))  \nonumber \\
&=& \int {\rm d}s \; P(\bar{w}(s)) J(s) g^{\gamma}_{rs} (\bar{w}(s)-\bar{w}(r))
          \nonumber \\
& & \;\;
   +  \lambda \cdot  P(\bar{w}(r)) J(r)
   \cdot \int {\rm d}r^{'} \; g^{\gamma}_{r^{'}r} (\bar{w}(r^{'})-\bar{w}(r)).
\end{eqnarray}
The derivation can be performed analoguous to 
(Ritter, Martinetz
\&    
Schulten 1992).
In the continuum limit there is always an exactly matching winning weight
vector $\bar{w}_s=v$.
Further the integration variable is substituted,
${\rm d}v={\rm d}\bar{w}_s=J(s){\rm d}s$,
and we define the abbreviation $\bar{P}:=P(\bar{w}(r))$.
In the first integrand  $\bar{P}J$ has to be expanded in powers of $q:=s-r$.
Within the second integral $\bar{P}J$ is evaluated only at $r$.
Thus the integration yields in leading order in $q$:
\begin{eqnarray}
0 &=& \gamma^2 \left(\frac{{\rm d}\bar{w}}{{\rm d}r}
           \frac{{\rm d}(\bar{P}J)}{{\rm d}r}
           + \frac{1}{2} \bar{P}J \frac{{\rm d}^2\bar{w}}{{\rm d}r^2}\right)
  \nonumber\\ & &    
  + \lambda \cdot \bar{P}J \cdot \int {\rm d}q \; g^{\gamma}_{0q}
      \; (\!\!\!\!\!\underbrace{q \frac{{\rm d}\bar{w}}{{\rm d}r}}_{\mbox{\tiny
contribution 0}}\!\!\!\!\!
       + \frac{q^2}{2} \frac{{\rm d}^2\bar{w}}{{\rm d}r^2})    \nonumber \\
0 &=& \gamma^2 \left(J \frac{{\rm d}(\bar{P}J)}{{\rm d}r}
           + \frac{\bar{P}J}{2} \frac{{\rm d}J}{{\rm d}r}
      + \lambda \frac{\bar{P}J}{2} \frac{{\rm d}J}{{\rm d}r}\right).
\end{eqnarray}
Further we have to require
 $\gamma\neq 0,$    $P\neq 0,$   $d\bar{P}/dr\neq 0$.
Then the ansatz of an {\sl universal local magnification law}
$J(r)=J(\bar{P}(r))$, 
\mbox{i.e.} $J$ depends only on the {\sl local} value of $P$,
that may be expected for the one-dimensional case only,
requires $J$ to fulfill the differential equation 
\begin{eqnarray}
0 &=& \frac{J}{\bar{P}}
+ (1 + \frac{1}{2} + \frac{\lambda}{2}) \frac{{\rm d}J}{{\rm d}\bar{P}}
\end{eqnarray}
or
\begin{eqnarray}
\frac{{\rm d}J}{{\rm d}\bar{P}} &=& - \frac{2}{3+\lambda} \frac{J}{\bar{P}}.
\end{eqnarray}
It has a power law solution,
(provided that $\lambda\neq{-3}$),
which verifies the ansatz made above, $J$ being a function of the
local density only,
\begin{eqnarray}
M=\frac{1}{J} \sim P(v)^{\mbox{${\frac{2}{3+\lambda}}$}}.
\label{eq_wrkmagni}
\end{eqnarray}
Thus the magnification exponent is given by
$\frac{2}{3+\lambda}$ and can be tuned from $1/2$ to 1 
(see Fig.~\ref{fig:wrktheo})
within the range of stability.

For the $\lambda=1/2$ choice of the Winner-Relaxing Kohonen Algorithm
the magnification factor follows an exact power law with
magnification exponent $\rho=4/7$, which is smaller than
$\rho=2/3$ 
for the classical Self Organizing Feature Map
(Ritter \&  Schulten 1986),
 but is still much 
larger than $\rho=1/3$ for Vector Quantization and Neural Gas.
In any case, the maps resulting from
the choices $\lambda=1/2$ and $\lambda=0$
are not optimal in terms of information theory.

\section{Enhancing the Magnification}
From this result one would try to invert the
Relaxing Effect by choice of negative
values for $\lambda$, which means to ``enforce'' the winner. 
In fact, the choice of $\lambda=-1$ leads to
the magnification exponent 1.

The magnification law (\ref{eq_wrkmagni}) 
is verified numerically as is shown in
Table~\ref{wrtab}.
Apart from the fact that the exponent can be varied by 
{\it a~priori} parameter choice between $1/2$ and $1$, the
simulations show that 
our Winner Relaxing Algorithm
is able to establish information-theoretically
optimal self-organizing maps
in the ``winner enforcing'' case ($\lambda<0$).

\clearpage
\begin{table}[htbp]
\newcommand{\myklein}{\footnotesize\scriptsize$\pm$}
\vspace*{-0.4cm}
\begin{center}
\begin{tabular}[t]{|l|llll|l|llll|}
\hline
$\!\!\downarrow\!\gamma$~$\lambda\!\rightarrow\!\!\!$ 
 & -1& -3/4 &-1/2 &-1/4  &0  &1/4 &1/2 &3/4 &1 \\ 
\hline$\!$0.1 
& 0.29 & 0.29 & 0.23 & 0.29 & 0.27 & 0.25 & 0.26 & 0.27 & 0.27$\!$\\
&\myklein{}.04 &\myklein{}.04 &\myklein{}.04 & \myklein{}.04
&\myklein{}.05
&\myklein{}.04 &\myklein{}.04 &\myklein{}.04 & \myklein{}.05\\
\hline$\!$0.5 
& 0.49 & 0.46 & 0.43 & 0.45 & 0.43 & 0.40 & 0.39 & 0.37 & 0.34$\!$\\
&\myklein{}.02 &\myklein{}.01 &\myklein{}.02 & \myklein{}.01
&\myklein{}.01
&\myklein{}.01 &\myklein{}.02 &\myklein{}.01 & \myklein{}.01\\
\hline$\!$1.0 
& 0.75 & 0.77 & 0.68 & 0.67 & 0.61 & 0.58 & 0.58 & 0.53 & 0.51$\!$\\
&\myklein{}.04 &\myklein{}.02 &\myklein{}.02 & \myklein{}.02
&\myklein{}.01
&\myklein{}.01 &\myklein{}.01 &\myklein{}.01 & \myklein{}.01\\
\hline$\!$2.0 
& 0.93 & 0.86 & 0.77 & 0.71 & 0.65 
& 0.61 & 0.57 & 0.53 & 0.50$\!$\\
&\myklein{}.03 &\myklein{}.02 &\myklein{}.02 & \myklein{}.01
&\myklein{}.01
&\myklein{}.01 &\myklein{}.01 &\myklein{}.01 & \myklein{}.01\\
\hline$\!$5.0 
& 0.99 & 0.88 & 0.80 & 0.72 & 0.66 & 0.61 & 0.57 & 0.53 & 0.50$\!$\\
&\myklein{}.05 &\myklein{}.04 &\myklein{}.03 & \myklein{}.02
&\myklein{}.02
&\myklein{}.02 &\myklein{}.02 &\myklein{}.02 & \myklein{}.02
\normalsize\\
\hline$\!$Theory:$\!\!\!$ 
& 1.00 & 0.89 & 0.80 & 0.73 & 0.67 & 0.62 & 0.57 & 0.53 & 0.50$\!$\\
\hline
\end{tabular}
\end{center}
\vspace*{-2ex}
\caption[Numerical results of WRK]{\small\sf
Magnification exponent of the Winner Relaxing Algorithms
determined numerically from a sample setup with
$200$ neurons and $2\cdot{}10^7$ update steps and a
 learning rate of $0.1$.
The input space was the unit interval,
the stimulus probability density
was chosen exponentially as $\exp(-\beta{}w)$ with $\beta=4$.
After an adaptation period of $5\cdot{}10^7$ learning steps
 further $10\%$ of learning steps
were used to calculate average slope and its fluctuation
of $\log{}J$ as a function of $\log{}P.$ (The first and last $10\%$
of neurons were excluded to eliminate boundary effects).
The small numbers denote the fluctuation of the exponent
through the final $10\%$ of the experiment.
For small $\gamma$, the 
neighborhood interaction becomes too weak. If the Gaussian 
neighborhood extends over some neurons ($\gamma=5$),
the exponent follows the predicted dependence of $\gamma$
given by $2/(3+\lambda)$. For $|\lambda|>1$ the system
is instable, this is the case where the additional update
term of the winner is larger than the sum over all other update
terms in the whole network.
Tuning of the parameter $\mu$ did not seem to extend the
region of stability.
As the relaxing effect 
is inverted for
$\lambda<0$, fluctuations are larger than 
in the Kohonen case.
\label{wrtab}}
\end{table}
\vspace*{-5ex}
\section{Ordering time and Stability Region}
At least for the 2D$\to$2D case, 
the Winner-Relaxing Kohonen Algorithm
was reported as `somewhat faster' (Kohonen 1991)
in the initial
ordering process.
In a 1D$\to$1D sample setup
(Claussen 2003),
a marginally quicker ordering was 
observed for negative $\lambda$, 
at least at a relatively high learning rate
$\eta=1$.
As a lot of parameters 
and the input distribution itself 
influence the ordering time and decay of
fluctuations, different results may be 
obtained;
e.g., a small fraction of input distributions
containing topological kinks take much longer
to become ordererd,
thus minimal, maximal, averaged, and inverse averaged
ordering time will deviate.

\begin{figure}[bhpt]
\centerline{\epsfig{file=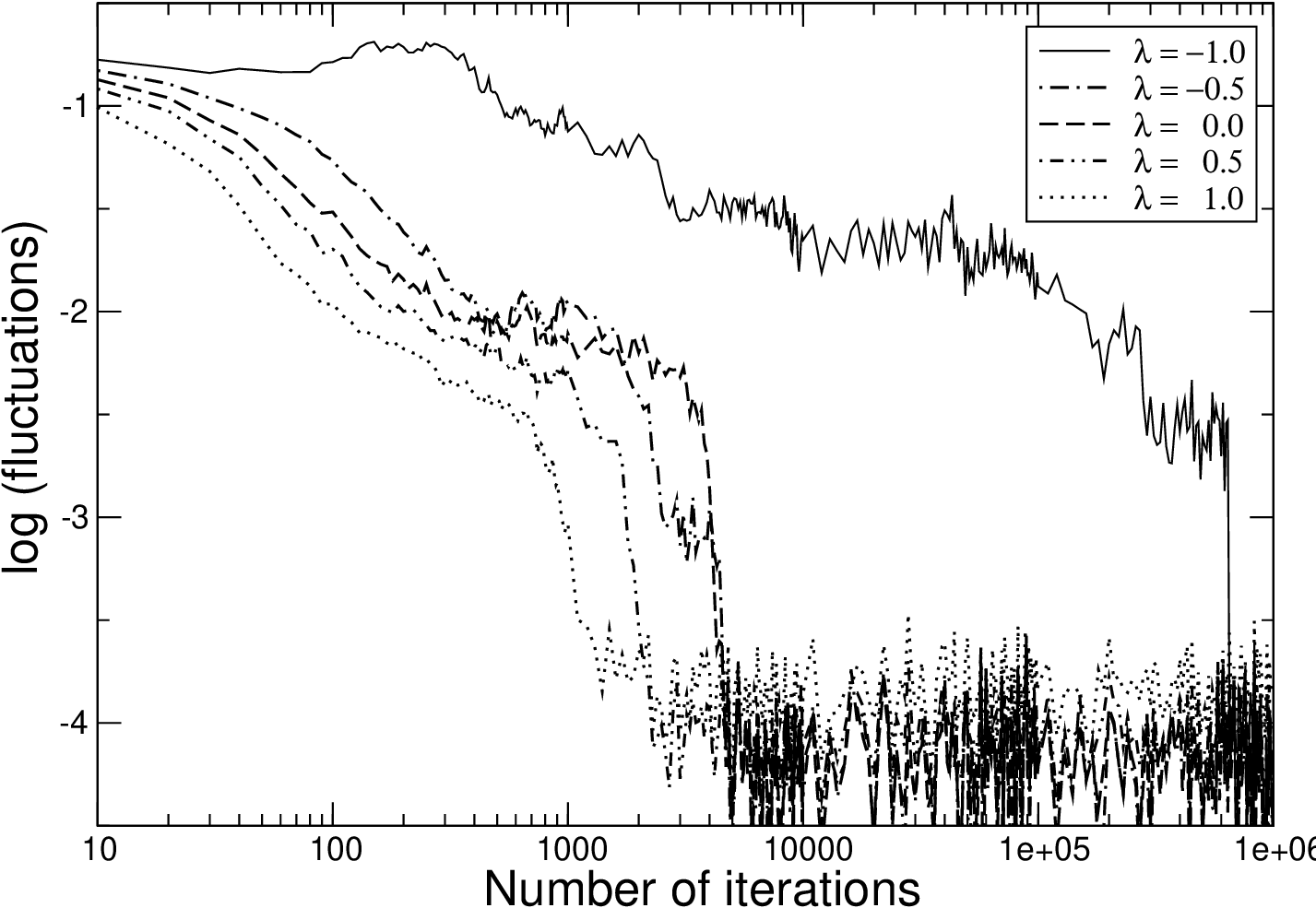,width=0.85\columnwidth}}
\caption{\small\sf
Time dependence (every tenth iterate shown) 
of the log rms fluctuations 
for different
$\lambda$.
Here the same setup of a single run with $\gamma=1.0$,
$\eta=0.1$
and 10 neurons is being used;  each run starts with
the same configuration and random initial values 
between 0 and 1.
For $\lambda>0$
a quicker ordering is observed.
\label{fig_conv}
}
\vspace{.9ex}
\centerline{
\epsfig{file=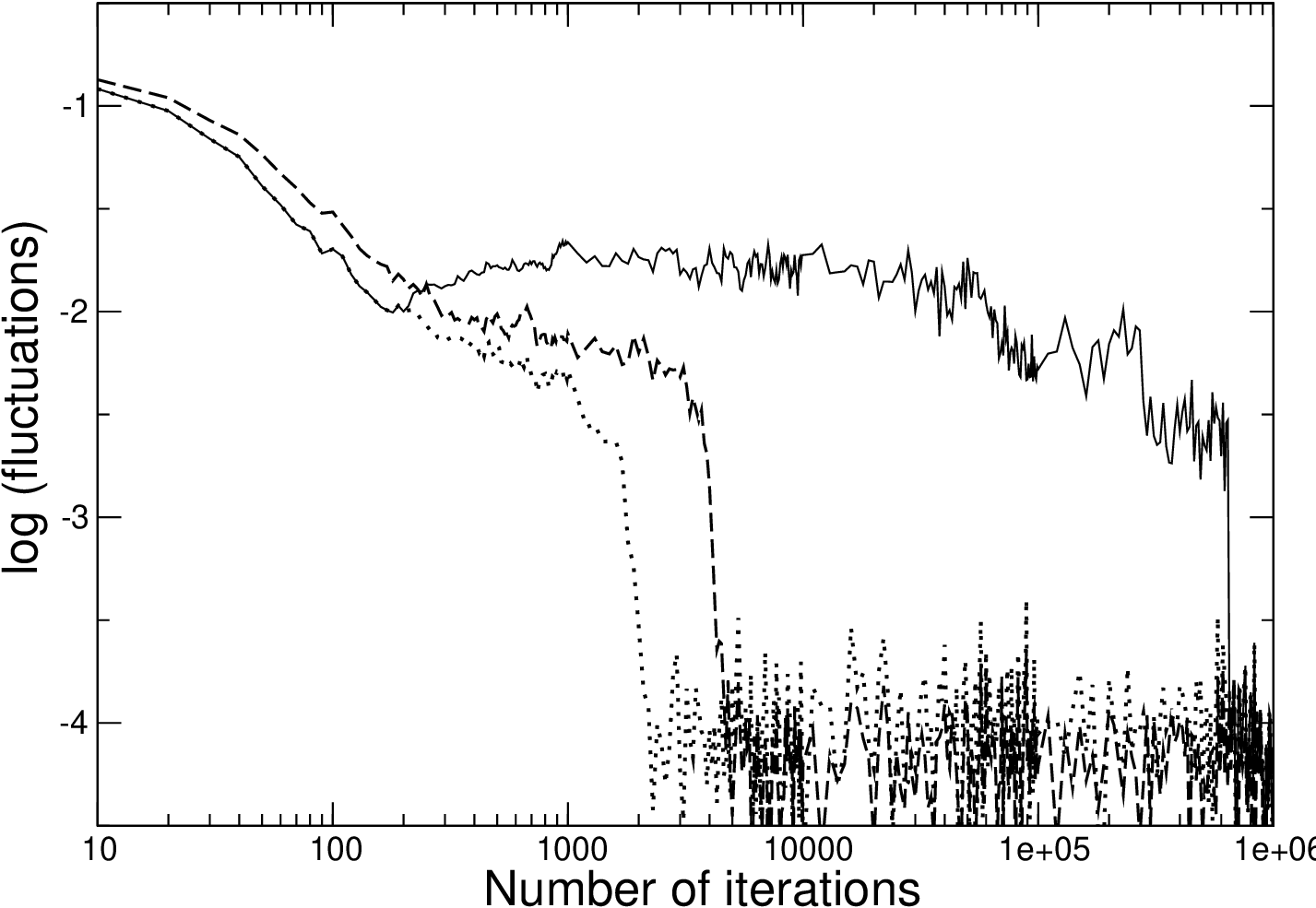,width=0.85\columnwidth}
}
\caption{\small\sf
Fast learning using a 
simple 
switching strategy. 
Starting with $\lambda=1/2$, 
ordering is acheived quickly.
At iteration step 2000, $\lambda$ is immediately changed to -1
(dotted).
This speeds up the learning phase by two orders of magnitude
compared to starting with $\lambda=-1$, and by a factor 4 
compared to $\lambda=0$ (dashed, shown for comparison).
If the duration of the initial ordering phase is
underestimated, again a long learning phase results
(solid line; switch at step 200).
\label{fig_switch}
}
\end{figure}

If one instead investigates the time dependence of the fluctuations,
for positive values of $\lambda$ a considerably quicker
decay is observed (Fig.~\ref{fig_conv}), being consistent 
with the observation by Kohonen (1991) mentioned above.

\clearpage
\clearpage
\noindent
These simulations indicate that  
for obtaining optimal 
magnification, the price of a longer learning phase may have
to be paid.
However, this drawback can be circumvented
by combining the advantages of both
$\lambda$ ranges;
i.e.\
using $\lambda>1$
in the initial phase to speed up ordering, 
and switching to $\lambda=-1$
after a considerable decay of fluctuations
(Fig.~\ref{fig_switch}).
No complicated time-dependence of this parameter switch
has been used, and 
neither learning rate nor neighborhood have been 
changed during the simulation.

\begin{figure}[pbht]
\vspace*{-1ex}
\centerline{\epsfig{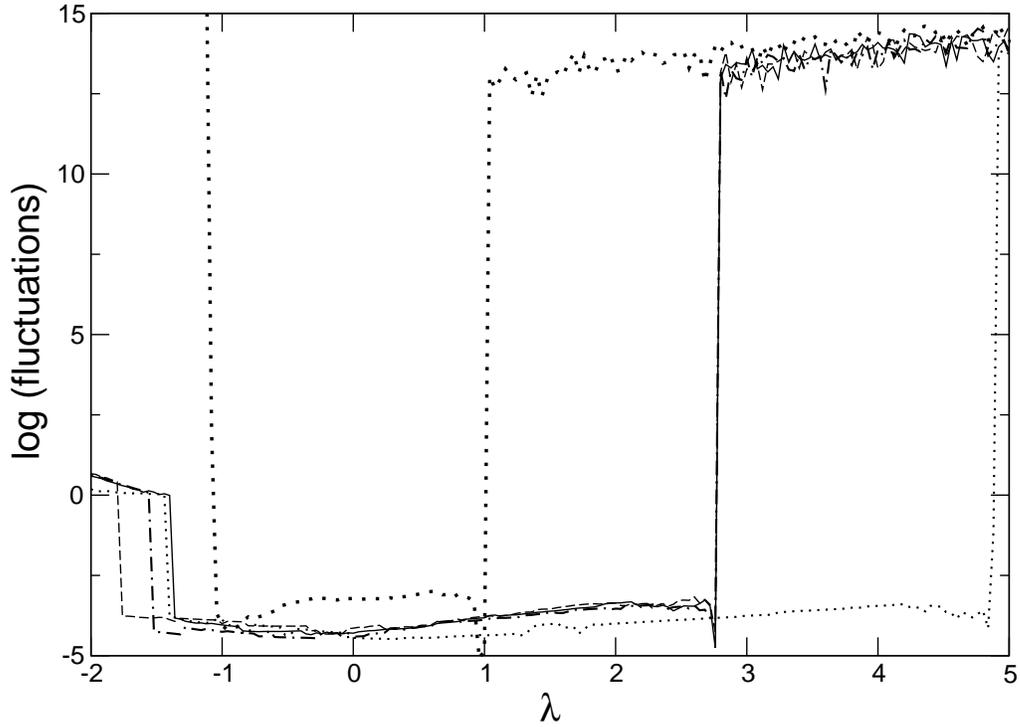}}
\caption{\small\sf 
For $\mu\in[-1,+1]$ the common
 stability range is $\lambda\in[-1,+1]$.
For $\lambda<-1$,
the log rms of the weight vector 
differences $w_r-w_{r-1}$ diverges,
but extremely long quiet transients are observed there.
In the upper range $\lambda>+1$, 
making use  of the diagonal term by using 
$\mu\neq{}0$ extends the stability range.
The plots correspond to $10^7$ (straight), 
$10^6$ (dash-dotteded),
and $10^5$ (dashed),respectively, for $\mu=0$.
For $10^7$ iterations, also the cases 
$\mu=-1$ (thin dots) 
and $\mu=+1$ (thick dots) are shown.
Parameters are $\gamma=1.0$, $\eta=0.1$, and 10 neurons 
are initialized near an equidistant chain with
noise of amplitude 0.01 added.
\label{fig_stab}}
\end{figure}

The last important issue to be addressed 
is the dependence
of stability on the parameter $\lambda$, especially
at the border $-1$.
Fortunately, the algorithm appears to be 
stable (in the 1D$\to$1D case) in the whole
range $-1\leq\lambda\leq+1$, 
as shown in Fig.~\ref{fig_stab}.
On both borders the Winner Relaxing learning
remains stable.
Thus, the full range of magnification exponents 
between $1/2$ and $1$ can be acheived.

\begin{figure}[htbp]
\centerline{\epsfig{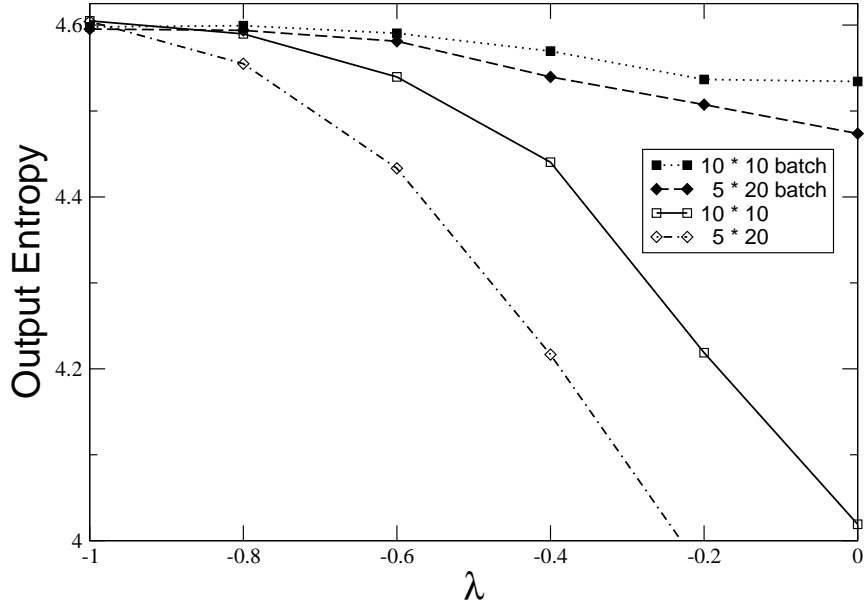}}
\caption{\small\sf
Entropy enhancement for the 2D$\to$2D case
for network geometries of $10*10$
and $5*20$ neurons.
The data density was $\sin({\pi}v_1)\cdot\sin({\pi}v_2)$ 
within the unit square,
$\gamma=5.0$, and
$\eta$ was decreased from $0.01$ to $0.001$
during $10^6$ learning steps.
Alternatively, batch learning (over 100 steps) 
has been used; here $\eta$ was decreased from $0.05$ to $0.001$,
$\gamma=2.0$
in the first $2\cdot10^5$ steps ordinary SOM 
learning was applied ($\gamma=5.0$, $\lambda=0$).
In all cases, for
$\lambda=-1$ the entropy is
enlarged compared to the unmodified case
$\lambda=0$,
and close to the optimum ($\ln 100=4.605$).
\label{fig:compare}}
\end{figure}

\clearpage
In higher dimensions no universal magnification law
is expected, but one can evaluate the output entropy
for a given input distribution and network.
As shown in Fig.~\ref{fig:compare},
the enhancement of output entropy by 
Winner Relaxing learning 
is effective also in the twodimensional case,
where however parameters have to be chosen
more carefully.

\section{Discussion}
After our first study
(Claussen 1992),
Herrmann et al.\ (1995)
introduced annother modification of the
learning process, which was also applied 
 to the Neural Gas algorithm
(Villmann \& Herrmann 1998).
Their central idea is to use a 
learning rate $\eta$ being
locally dependent on the input probability density 
and also an exponent 1 can be obtained. 
As the input probability density should not be available
to a neural map that self-organizes 
from stimuli drawn from that distribution,
it is estimated from the actual local reconstruction
mismatch (being an estimate for the size of the
Voronoi cell) and from the time elapsed since
the last time being the winner.
Both operations require additional memory and
computation, and, 
due to the estimating character, the learning rate
has to be bounded in practical use.
This {\sl localized learning} was overall easier
applicable and overcame the stability problems
of the early approach of {\sl conscience learning}
(deSieno 1988).

Another systematic method,
the extended Maximum Entropy Learning Rule, 
has been introduced by
van Hulle (1997).
It approximates a map of maximal output entropy
for arbitrary dimension, alhough in higher 
dimensions the handling of the 
quantization regions becomes less practial
(van Hulle 1998).
A quite different approach being also capable of generating
equiprobabilistic maps is via kernel optimization
(van Hulle 1998, 2000, 2002),
i.e.\  neighborhood kernel radii 
themselves become learning parameters,
in addition to the weight vectors defining the kernel centers.
Other approaches, also influencing
magnification, 
consider the {\sl selection} of the winner
to be probabilistic, leading to  elegant
statistical approaches  to  potential functions, 
as given by
Graepel et al.\ (1997)
and 
Heskes (1999).

As shown recently (Claussen \& Villmann 2004), 
the Winner Relaxing concept can also be transferred
successfully to the Neural Gas,
confirming the utility of this class of
learning rules.

\section{Conclusions}
The Linsker, Elastic Net 
and Winner-Relaxing Kohonen algorithms
can be derived from an extremal principle, given by information theory,
physical motivations, and reconstruction error, respectively.
In this paper we have chosen the magnification law
to indicate how close the algorithm reaches the adaptation
properties of a map of maximal mutual information.
The magnification law is one quantitative property that
both is accessible by neurobiological experiments
and manifests as a quantitative control parameter of a neural map 
used as vector quantizer in applications.
A map of maximal mutual information uses all neurons
with same probability, i.e.\ their firing rate will be
equal. 

In this work we have investigated the Winner Relaxing approach
to establish a new family of vector quantizers.
The shift from Kohonen ($\rho=2/3$) to Winner Relaxing
Kohonen algorithm ($\rho=4/7$) seems to be marginal,
if the emphasis is laid on the existence of a potential function.
If a large magnification exponent is desired,
the Winner Relaxing Kohonen Algorithm (with $\lambda=-1$)
combines simple computation with
a magnification corresponding to maximal mutual information.

\clearpage\noindent
{\bf Acknowledgements:}
The author wants to thank H.G.\ Schuster for
raising attention to the topic and for stimulating 
discussions.

\section*{References}
\begin{description}

\item[]
Claussen, J.C.\  (born Gruel) (1992).
{\it Selbstorganisation Neuronaler Kar\-ten.} 
Diploma thesis, 
Kiel, Germany.
\\[-4.5ex]

\item[]
Claussen, J.C.\  \& Schuster, H.G.\
(2002).
 Asymptotic level density of the Elastic Net Self-Organizing Map, 
Proc.\ ICANN 2002.
\\[-4.5ex]

\item[]
Claussen, J.C.\ (2003). 
Winner Relaxing and Winner-Enhancing Kohonen Maps: Maximal
Mutual information from Enhancing the Winner.
{\it Complexity}, 
{8(4)}, 15-22. 
\\[-4.5ex]

\item[]
Claussen, J.C.\  \& Villmann, T.\
(2004).
Magnification Control in Winner Relaxing Neural Gas.
{\it Neurocomputing}, 
in print.
\\[-4.5ex]

\item[]
Dersch, D.R., \& Tavan, P.\ (1995). 
Asymptotic level density in topological feature maps. 
{\it IEEE Transactions on Neural Networks},  6, 230-236.
\\[-4.5ex]

\item[]
DeSieno, D.\ (1988).
Adding a conscience to competitive learning.
In: {\em Proc. ICNN'88, International Conference on Neural
Networks},
  pp.\ 117-124, IEEE Service Center, Piscataway, NJ.
\\[-4.5ex]

\item[]
Durbin, R.\ \& Willshaw, D.\ (1987). 
An analogue approach to the Travelling Salesman Problem using an Elastic Net Method. 
{\it Nature}, {326}, 689-691.
\\[-4.5ex]

\item[]
Herrmann, M., Bauer, H.-U., Der, R.\ (1995).
Optimal Magnification Factors in Self-Organizing Maps.
Proc. ICANN 1995,
pp. 75-80.
\\[-4.5ex]

\item[]
Graepel, T., Burger, M., \& Obermayer, K.\
(1997).
Phase transitions in stochastic self-organizing maps.
{\it Phys.\ Rev.\ E,} 56, 3876-3890.
\\[-4.5ex]

\item[]
Heskes, T.\ (1999).
Energy functions for self-organizing maps.
In: Oja and Karski, ed.: Kohonen Maps, Elsevier.
\\[-4.5ex]

\item[]
van Hulle, M.\ M.\ (1997).
          Nonparametric density estimation
          and regression achieved with
          topographic maps maximizing the
          information-theoretic entropy of their
          outputs.
{\it Biological Cybernetics},
77, 49-61.
\\[-4.5ex]

\item[]
van Hulle, M.\ M.\ (1998).
Kernel-based equiprobabilistic topographic map formation.
{\it Neural
Computation}. 
10,
1847-1871. 
\\[-4.5ex]

\item[]
van Hulle, M.\ M.\ (2000).
{\it 
Faithful Representations and Topographic Maps.}
Wiley.
\\[-4.5ex]

\item[]
van Hulle, M.\ M.\ (2002).
Kernel-based topographic map formation by local density modeling.
{\it Neural
Computation,}       
14,
1561-1573.
\\[-4.5ex]

\item[]
Kohonen, T.\ (1982).
Self-Organized Formation of Toplogically Correct Feature Maps.
{\it Biological Cybernetics}, {43}, 59-69.
\\[-4.5ex]

\item[]
Kohonen, T.\ (1991).
Self-Organizing Maps: Optimization Approaches.
In: {\it Artificial Neural Networks}, 
ed. T.\ Kohonen et~al. {(North-Holland, Amsterdam)}.
\\[-4.5ex]

\item[]
Linsker, R.\ (1989).
How To generate Ordered maps by Maximizing the Mutual Information
between Input and Output Signals.
{\it Neural Computation}, {1}, 402-411.
\\[-4.5ex]

\item[]
Obermayer, K., Blasdel, G.G., \&  Schulten, K.\ (1992). 
Statistical-mechanical analysis of self-organization and pattern formation
during the development of visual maps.
{\it Phys. Rev. A}, {45}, 7568-7589.
\\[-4.5ex]

\item[]
Plumbley, M.D.\ (1999).
Do cortical maps adapt to optimize information density?,
{\sl Network}, 10, 41-58.
\\[-4.5ex]

\item[]
Ritter, H.\ \&  Schulten, K.\ (1986).
On the Stationary State of Kohonen's Self-Organizing Sensory 
Mapping. 
{\it Biol.\ Cybernetics}, {54}, 99-106.
\\[-4.5ex]

\item[]
Ritter, H.\ \&  Schulten, K.\ (1988).
Convergence Properties of Kohonen's Topology Conserving Maps:
Fluctuations, Stability and Dimension Selection.
{\it Biological Cybernetics}, {60}, 59-71.
\\[-4.5ex]

\item[]
Ritter, H.\ (1991). 
Asymptotic Level Density for a Class of Vector
Quantization Processes. 
{\it IEEE Trans.\  Neural Networks},
{2}, 173-175.
\\[-4.5ex]

\item[]
Ritter, H., Martinetz, T., \& Schulten, K.\  (1992).
Neural Computation and Self-Organizing Maps: An Introduction.
Addison-Wesley, 
NY.
\\[-4.5ex]

\item[]
Simic, P.D.\ (1990). 
Statistical mechanics as the underlying theory 
of `elastic' and `neural' optimizations. 
{\it Network}, {1}, 89-103.
\\[-4.5ex]

\item[]
Villmann, T., \& Herrmann, M.\ (1998).
Magnification Control in Neural Maps,
Proc. ESANN 1998, pp. 191-196.
\\[-4.5ex]


\end{description}
\underline{~~~~~~~~~~~~~~~~~~~~~~~~~~~~~~~~}\\
\small \footnotesize Manuscript submitted August 15, 2002; 
revised June 4 \& September 24, 2004; accepted November 1, 2004.

\end{document}